\documentclass[aps,prd,twocolumn,superscriptaddress,showpacs,floatfix,preprintnumbers]{revtex4-2}

\usepackage[dvipsnames]{xcolor}     
\definecolor{lcolor}{rgb}{0.5,0,0}
\definecolor{citcolor}{rgb}{0,0,1}
\usepackage[breaklinks,colorlinks,urlcolor=blue,citecolor=citcolor,linkcolor=lcolor,linktoc=all]{hyperref}
\usepackage{color}
\usepackage{graphicx}	
\graphicspath{{./figures/}}
\usepackage[utf8]{inputenc}
\usepackage{amsmath} 
\usepackage{amssymb}
\usepackage{slashed}
\usepackage{bm,bbm,bbold}
\usepackage{booktabs}
\usepackage{dcolumn}
\usepackage{dsfont}
\usepackage{MnSymbol} 
\usepackage{cancel}
\usepackage{booktabs}

\DeclareMathOperator*{\SumInt}{%
\mathchoice%
  {\ooalign{$\displaystyle\sum$\cr\hidewidth$\displaystyle\int$\hidewidth\cr}}
  {\ooalign{\raisebox{.14\height}{\scalebox{.7}{$\textstyle\sum$}}\cr\hidewidth$\textstyle\int$\hidewidth\cr}}
  {\ooalign{\raisebox{.2\height}{\scalebox{.6}{$\scriptstyle\sum$}}\cr$\scriptstyle\int$\cr}}
  {\ooalign{\raisebox{.2\height}{\scalebox{.6}{$\scriptstyle\sum$}}\cr$\scriptstyle\int$\cr}}
}

\makeatletter
\def\@fnsymbol#1{\ensuremath{\ifcase#1\or *\or \dagger\or \ddagger\or
   \mathsection\or \mathparagraph\or \|\or **\or \dagger\dagger
   \or \ddagger\ddagger \else\@ctrerr\fi}}
\makeatother

\allowdisplaybreaks

\definecolor{color0}{HTML}{648FFF} 
\definecolor{color1}{HTML}{785EF0} 
\definecolor{color2}{HTML}{DC267F} 
\definecolor{color3}{HTML}{FE6100} 
\definecolor{color4}{HTML}{FFB000} 

\makeatletter
\g@addto@macro\bfseries{\boldmath}
\makeatother

\usepackage{tikz, tikzscale}
\usetikzlibrary{calc, fadings, shadings, decorations.pathmorphing, decorations.pathreplacing, decorations.shapes}
\usepackage[customcolors]{hf-tikz}
\usepackage{mciteplus}

\makeatletter 

\renewcommand\onecolumngrid{
\do@columngrid{one}{\@ne}%
\def\set@footnotewidth{\onecolumngrid}
\def\footnoterule{\kern-6pt\hrule width 1.5in\kern6pt}%
}

\renewcommand\twocolumngrid{
        \def\footnoterule{
        \dimen@\skip\footins\divide\dimen@\thr@@
        \kern-\dimen@\hrule width.5in\kern\dimen@}
        \do@columngrid{mlt}{\tw@}
}%

\makeatother    

\newcommand{\pqcd}{p}
\newcommand{\ppq}{p_\mathrm{PQ}}
\newcommand{\mub}{\mu_\mathrm{B}}
\newcommand{\muq}{\mu_\mathrm{q}}
\DeclareMathOperator{\tr}{tr}

\begin{document}

\title{\texorpdfstring{A new approach to determine the thermodynamics of deconfined matter \\ to high accuracy}{A new approach to determine the thermodynamics of deconfined matter to high accuracy}}

\preprint{HIP-2025-32/TH}

\author{Tyler Gorda}
\email{gorda.1@osu.edu}
\affiliation{Center for Cosmology and AstroParticle Physics (CCAPP), Ohio State University, Columbus, OH 43210}
\affiliation{Department of Physics, The Ohio State University, Columbus, OH 43210, USA}
\author{Pablo Navarrete}
\email{pablo.navarrete@helsinki.fi}
\affiliation{Department of Physics and Helsinki Institute of Physics,
P.O.~Box 64, FI-00014 University of Helsinki, Finland}
\author{Risto Paatelainen}
\email{risto.paatelainen@helsinki.fi}
\affiliation{Department of Physics and Helsinki Institute of Physics,
P.O.~Box 64, FI-00014 University of Helsinki, Finland}
\affiliation{Department of Physics and Astronomy, FI-20014 University of Turku, Finland}
\author{Leon Sandbote}
\email{leon.sandbote@helsinki.fi}
\affiliation{Department of Physics and Helsinki Institute of Physics,
P.O.~Box 64, FI-00014 University of Helsinki, Finland}
\author{Kaapo Seppänen}
\email{kaapo.seppaenen@unibe.ch}
\affiliation{Department of Physics and Helsinki Institute of Physics,
P.O.~Box 64, FI-00014 University of Helsinki, Finland}
\affiliation{AEC, Institute for Theoretical Physics, University of Bern, Sidlerstrasse 5, CH-3012 Bern, Switzerland}

\begin{abstract}
We demonstrate that at finite density and sufficiently high temperatures, phase-quenched (PQ) lattice simulations combined with perturbation theory provide a new precision approach to determining the thermodynamics of QCD across a wide arc of the phase diagram where the strong coupling constant $\alpha_s$ remains small. In this regime, nonperturbative pairing effects in the PQ theory are parametrically suppressed, so that the difference between the PQ and full QCD pressures becomes a small perturbative correction. We compute this correction up to and including $O(\alpha_s^{7/2})$ using electrostatic QCD together with a novel numerical method to compute four-loop sum-integrals. This enables the determination of the perturbative QCD pressure with precision beyond the current state of the art while including nonperturbative pure-gluonic contributions from the lattice.
\end{abstract}

\maketitle

\section{Introduction}

The bulk thermodynamics of Quantum Chromodynamics (QCD) across its phase diagram still lack a detailed quantitative understanding, despite decades of theoretical and experimental effort \cite{Gross:1973id,Politzer:1973fx,Gross:2022hyw}.
This is due to a lack of both first-principles calculations and experimental data throughout a large portion of the temperature $T$ and baryon chemical potential $\mub$ plane.
At high $T$ and low $\mub$, lattice Monte Carlo simulations \cite{Aoki:2006we} have established a crossover from hadronic matter to a quark--gluon plasma, also probed in heavy-ion collisions  \cite{Gardim:2019xjs}.
However, these methods break down for $\mub/T \gtrsim 3.5$ \cite{HotQCD:2018pds,Borsanyi:2020fev,Abuali:2025tbd} due to the sign problem \cite{deForcrand:2009zkb,Nagata:2021ugx}, while experimental data does not currently probe larger $\mub$ \cite{Alba:2014eba,Becattini:2016xct,STAR:2017sal,Andronic:2017pug}.
Hence, in this regime, alternative techniques such as functional methods \cite{Fu:2019hdw,Gao:2020fbl,Gunkel:2021oya}, holography \cite{Jarvinen:2021jbd,Hoyos:2021uff,Kovensky:2021ddl,Demircik:2021zll,Ecker:2025vnb}, or approaches combining multiple inputs \cite{Parotto:2018pwx,Noronha-Hostler:2019ayj,An:2021wof,MUSES:2023hyz,Mroczek:2024sfp} must be employed.

As a result, first-principles information about QCD thermodynamics, encoded in the equation of state (EoS), is limited to disconnected regions of the phase diagram, leaving the intermediate-density, finite-temperature regime essentially unconstrained. This region is, however, phenomenologically highly relevant for forthcoming heavy-ion experiments at intermediate beam energies, including low-energy scan programs at the Facility for Antiproton and Ion Research, and for binary neutron-star mergers, both of which crucially depend on the EoS at finite $T$ and $\mub$ for $N_f=3$ quark flavors.

At very large values of $T$ or $\mub$, the strong coupling constant $\alpha_s$ becomes small, and perturbative calculations directly within QCD itself become feasible.
These calculations have reached the $O(\alpha_s^{5/2})$ level throughout the phase diagram \cite{Kurkela:2016was}, including the effects of a nonzero strange-quark mass \cite{Gorda:2021gha}, by exploiting effective field theories of long-wavelength screened gluonic fluctuations \cite{Braaten:1989mz,Manuel:1995td,Braaten:1995cm} appearing in loop corrections.
Though these are first-principles calculations, they are not without challenges.
For example, it has long been known that perturbative QCD suffers from poor convergences in the high-$T$, low $\mub$ region \cite{Blaizot:2003iq}.
Additionally, the uncertainty quantification of errors from missing higher-order terms in the truncated perturbative series is currently rather limited \cite{Gorda:2023usm,Semposki:2024vnp,Semposki:2025etb}, with most works simply varying the renormalization scale in results by a factor of two to approximate this uncertainty.
Furthermore, at high $\mub$ and small $T$, gluonic attraction between quarks is expected to lead to the formation of a color superconductor, altering the ground state at high densities \cite{Barrois:1977xd,Bailin:1983bm,Alford:1997zt,Son:1998uk,Rapp:1997zu,Schafer:1999jg,Pisarski:1999tv,Alford:1998mk,Alford:2002kj,Alford:2007xm} and contributing a subdominant, nonperturbative pairing term to the pressure $p_\text{pairing} \propto \mub^2 \Delta^2$, with $\Delta$ the pairing gap.
There is ongoing discussion in the literature as to the size of these pairing corrections, with some recent works suggesting loop corrections to $p_\text{pairing}$ may be unexpectedly large \cite{Geissel:2025vnp}, while others suggest the leading-order (LO) coefficient may be smaller than previously computed \cite{Fujimoto:2025qwz, Abbott:2024vhj}.

\begin{figure}[t]

\hspace{-0.7cm}
\includegraphics[width=0.9\linewidth]{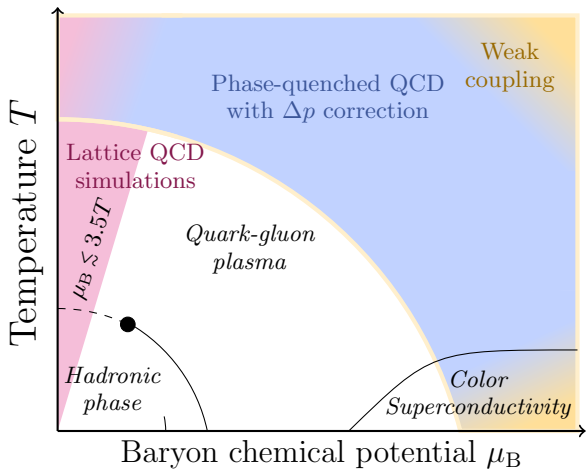}

\caption{The QCD phase diagram with its phases (italic text) and the regions accessible using various computational techniques (shaded regions and colored text). The PQ critical temperature of the paired phase restricts the applicability of PQ simulations to determine the QCD pressure, indicated by the blue-to-orange fade in the bottom right.}
\label{fig:QCDpD}
\end{figure}

Another approach to study the QCD phase diagram that has recently received renewed attention \cite{Moore:2023glb,Fujimoto:2023unl,Fujimoto:2023mvc} is to instead study the phase-quenched (PQ) theory, by dropping the complex fermion phases in the QCD path integral $Z$.
This removes the sign problem and allows for direct Monte Carlo lattice simulations.
In particular, the two-flavor PQ theory corresponds to QCD at nonzero isospin chemical potential $\mu_\mathrm{I} = \mu_u = -\mu_d$ with $u$ and $d$ denoting the up and down quarks~\cite{Cohen:2003ut}.
Since dropping the fermionic phase amounts to taking an absolute value of the fermionic determinant in $Z$, the pressure in the PQ theory must be greater than the true pressure $\pqcd$ and the two theories satisfy $\ppq(\{\mu_i\}) \geq \pqcd(\{\mu_i\})$ for \emph{any} number of fermions with a fixed set of chemical potentials $\{\mu_i\}$.
In Ref.~\cite{Moore:2023glb}, it was shown that $\Delta p := p_\text{PQ}-p$ is parametrically small of $O(\alpha_s^3)$ at exactly zero temperature and asymptotically large values of $\mub$---up to nonperturbative pairing corrections in the two theories---and that the LO contribution to the difference arises from a single \emph{finite}  Feynman diagram (up to fermionic orientations). 
The part of this ``Bugblatter'' diagram \cite{Blaizot:2001vr} contributing to this difference was subsequently computed numerically in Ref.~\cite{Navarrete:2024zgz}.

However, it has been demonstrated that nonperturbative pairing effects in the $N_f = 2$ PQ theory are strongly enhanced compared to QCD \cite{Son:2000xc, Fujimoto:2023mvc}, making it more difficult to extract useful information about QCD from PQ lattice simulations. Additionally, extrapolating simulations performed at large $\mub$ and finite $T$ to vanishing temperatures proves to be technically very challenging (for a discussion, see Ref.~\cite{Moore:2023glb}).

In this work, we show that these difficulties can be circumvented by considering higher temperatures and number of quark flavors, with $N_f=3$
being the most interesting scenario for phenomenological applications. 
Although phase quenching does not lead to a sensible physical theory in this case, it nevertheless has a well-defined pressure, and $\Delta p$ remains a rigorously finite and positive quantity that can be used to place rigorous constraints on the pressure of the physical theory through the relation $p = \ppq - \Delta p$.

Concentrating on the $\Delta p$ quantity, the arguments of Ref.~\cite{Moore:2023glb} demonstrating that the leading-order contribution arises from a single Feynman diagram extend to nonzero temperatures with only minor modifications. 
In particular, temperature-dependent infrared divergences can be systematically treated using well-known effective field theory techniques, while unknown nonperturbative quark-pairing effects are suppressed at sufficiently high temperatures, making lattice simulations more accessible by avoiding uncontrolled $T\to0$ extrapolations occurring at large $\mub$.
Therefore, combining PQ lattice simulations with a perturbative evaluation of $\Delta p$ enables an accurate determination of the thermodynamics of three-flavor QCD throughout a large arc in the phase diagram (see Fig.~\ref{fig:QCDpD}).
This information can then be used as a crucial boundary condition to constrain the EoS in the phenomenologically relevant region in a largely model-independent way (see e.g. Ref.~\cite{Komoltsev:2021jzg}).

In the following, we make this argument more precise and quantitative by estimating the critical temperature $T_c^\text{PQ}$ in the PQ theory, necessary to remove the unknown pairing corrections from $\Delta p$. 
Additionally, we compute the perturbative contributions to $\Delta p$ at $O(\alpha_s^3 \ln \alpha_s)$, $O(\alpha_s^3)$, and $O(\alpha_s^{7/2})$ for nonzero temperature and density. 
This provides a prescription for determining the QCD pressure across a large region of the phase diagram with an error of $O(\alpha_s^4)$,
while including nonperturbative pure-gluonic contributions from the lattice. This result surpasses the accuracy of state-of-the-art perturbative results \cite{Kurkela:2016was}.

\section{Review of PQ and the pairing gap}

The PQ theory can be constructed from the path integral of QCD with $N_f$ quark flavors,
\begin{equation}
    \label{eq:ZQCD}
    Z(T,\{\mu_i\}) = \int \mathcal{D} A_\nu \,e^{-S[A_\nu]} \prod_{i=1}^{N_f} \det( \mathcal{D}(\mu_i)),
\end{equation}
by replacing the fermion determinant by $|\det\mathcal{D}(\mu_i)| = \sqrt{\det\mathcal{D}(\mu_i) \det\mathcal{D}(-\mu_i)}$, a property following from $\gamma^5$-hermiticity.
Here $A_\nu$ are the gluonic fields, $S[A_\nu]$ is the usual gluonic action, $\{\mu_i\}$ is a set of $N_f$ quark chemical potentials, and ${\mathcal{D}(\mu_i) \coloneq \slashed{D}+m_i+\mu_i \gamma^0}$ is the Dirac operator.
By further using 
\begin{equation}
    \label{eq:half_fermion_sqrt}
    \sqrt{ \det \mathcal{D}(\pm\mu_i) } = \exp \left\{ \frac{1}{2} \cdot \tr \log \mathcal{D}(\pm\mu_i) \right\},
\end{equation}
one finds that the Feynman rules for the PQ theory are the same as in QCD with $2N_f$ ``half-fermions'', with each fermion loop receiving an additional symmetry factor of $1/2$. 
We denote the basis for these half-fermions $( q_+, q_-)^T$, where $q_+$ and $q_-$ are each made up of $N_f$ Dirac-spinors and the sign $\pm$ refers to the sign of the chemical potential of the quark.
For the rest of this work, we will specialize to the case of massless quarks with identical chemical potentials $\muq = \mub/3$.

To see how quantities differ from the case of isospin with $N_f = 2$, we now estimate the pairing gap in the PQ theory with $N_f = 3$ quark flavors.
For a more detailed discussion about this computation, see Appendix~\ref{appsec:PQpairing}.
The dominant contribution will be between quarks and antiquarks of the same color and opposing chemical potential, due to the exponential gap increase for color-neutral $q_+$--\ $\bar{q}_-$ pairing \cite{Son:2000xc,Fujimoto:2023mvc}. 
Moreover, instanton effects favor negative parity \cite{Son:2000xc}. 
We hence assume that the Dirac and color spaces of the half-fermion propagator decomposes into a tensor product, leaving one undetermined $N_f\times N_f$ matrix $\Phi$ parameterizing the flavor angles of the pairing. 
Without loss of generality, we can assume $\Phi = \Delta^\text{PQ} \mathds{1}_\mathrm{f}$ for the flavor angles.

Following the weak-coupling computations performed in Refs.~\cite{Pisarski:1999tv,Fujimoto:2023mvc} with an effective half-fermion action based on Ref.~\cite{Cornwall:1974vz}, we find the following estimate for the pairing gap at zero temperature:
\begin{equation}\label{eq:gap}
    \Delta^\text{PQ} = \muq \frac{16 \pi^{3/2}}{\alpha_s^{5/2}} \left( \frac{2}{N_f} \right)^{5/2} \exp\Bigg[{-\frac{3\pi^{3/2}}{4 \sqrt{\alpha_s}}- \frac{1}{4} - \frac{\pi^2}{16} }\Bigg].
\end{equation}
In the two-flavor case, this gap matches the isospin case of Ref.~\cite{Fujimoto:2023mvc}, and we see that for larger $N_f$ the gap is reduced in size. 
The $N_f$ dependence in this expression originates from the scaling $\propto N_f^{1/2}$ in the gluon screening mass, which modifies the gluonic propagation at low momentum and decreases the interaction strength between the quarks in the medium.
The expression for the gap in Eq.~\eqref{eq:gap} can also be related to the critical temperature \cite{Pisarski:1999tv}. 
For the pairing pattern discussed above, we find that the critical temperature $T_c^\text{PQ}$ is related to the gap as $T_c^\text{PQ} = e^{\gamma_\text{E}}  \Delta^\text{PQ}/\pi \approx 0.57 \Delta^\text{PQ}$,  reproducing the standard Bardeen--Cooper--Schrieffer pairing relation~\cite{Bardeen:1957mv}.

Using these relations, we can estimate the critical temperature for the $N_f =  3$ PQ theory assuming massless quark matter in equilibrium under flavor-changing weak processes (beta equilibrium) within full QCD.
Taking a reference value of $\alpha_s(2\,\mathrm{GeV}) = 0.2994$ in the $N_f=3$ case \cite{ParticleDataGroup:2008zun}, we then use three-loop running of the coupling to a lower scale $\bar{\Lambda} = X \sqrt{(2 \muq)^2 + (0.723 \times 4 \pi T)^2}$, with $X \in [1/2, 2]$ \cite{Kajantie:1997tt,Kajantie:2002wa,Kurkela:2009gj,Kurkela:2016was,Gorda:2021gha}. Using this procedure, we estimate the critical temperature for the $N_f = 3$ PQ theory to be about $T_c^\text{PQ} \simeq 40 - 110$~MeV for baryon chemical potentials $\mub  \simeq 2.5 - 3.0$~GeV.
For comparison, in the same range of baryon chemical potential, the critical temperature of the $N_f = 2$ PQ theory is enhanced, resulting in $T_c^\text{PQ} \simeq 110 - 340$~MeV assuming the value $\alpha_s(2\,\mathrm{GeV}) = 0.2710$ \cite{Aoki_2022,Fritzsch_2012}.
While these estimates are based on weak-coupling expansions potentially valid only at much higher chemical potentials, the fact that $T_c^\text{PQ}$ in the $N_f = 3$ case is decreased compared to the $N_f = 2$ case suggests that using three-flavor PQ lattice simulations to solve for the QCD pressure in the unpaired phase is possible throughout a large swath of the phase diagram.

\section{Difference between PQ and QCD at finite $T$ in the unpaired phase}

The corresponding Feynman rules for the PQ theory amount to averaging over the sign of the chemical potentials, thereby extracting the real part of the quark loop \cite{Navarrete:2024zgz}.
Hence, when computing $\Delta p$ perturbatively, the first contribution occurs with the four-loop Bugblatter diagrams, constructed from two quark loops attached by three gluon lines \cite{Moore:2023glb,Navarrete:2024zgz}, 
\begin{equation}
\label{eq:bugblatterdiags}
\raisebox{-0.45\height}{\includegraphics[height=1.4cm]{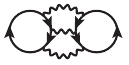}} +
\raisebox{-0.45\height}{\includegraphics[height=1.4cm]{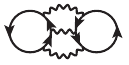}}.
\end{equation}
This remains true regardless of whether one has $T=0$ or $T>0$.
As further pointed out in Ref.~\cite{Moore:2023glb}, only the part of the quark loop that is symmetric in the color (latin) indices of the three incoming gluonic momenta---being proportional to the QCD color factor $d_{abc} \coloneq 2 \tr ( T_a \{T_b, T_c\})$---contributes to the difference $\Delta p$.

To evaluate $\Delta p$ at such high perturbative order, knowledge of the divergent properties of the diagrams in Eq.~\eqref{eq:bugblatterdiags} is crucial. 
Being a leading-order effect, on renormalization grounds one expects at least ultraviolet (UV) finiteness. 
Indeed, it was shown in Ref.~\cite{Navarrete:2024zgz} that, at $T=0$, this is achieved locally by summing up the two diagrams at the integrand level. 
For $T>0$, the vanishing of thermal effects in the UV limit guarantees that the same conclusion holds. 
That such internal cancellations occur in the most general case of nonzero temperatures (and even nonzero quark masses) is a feature that enables a numerical treatment of the full diagrams in Eq.~\eqref{eq:bugblatterdiags} without the need to construct any UV counterterms.

However, in medium, infrared (IR) sensitivities can arise from the dynamical screening of low-energy or low-momentum gluonic fluctuations, which are sensitive to the soft screening mass scale
\begin{equation}
\label{eq:screeningmass}
    m_\text{E}^2=\frac{g^2}{3}\left[ \left( N_c+\frac{N_f}{2} \right)T^2+\frac{3}{2\pi^2}N_f\muq^2 \right],
\end{equation}
with $N_c = 3$ the number of gluon colors and $g = (4 \pi \alpha_s)^{1/2}$ the gauge coupling.
As explained in Ref.~\cite{Kurkela:2016was}, two types of screening of ``soft'' low-momentum ($|\vec{p}| \lesssim m_\text{E}$) gluons occur at this scale. 
The first type refers to the screening of nonstatic modes of gluonic fields $A_\nu$, which can be described within the framework of Hard-Thermal/Dense-Loop (HTL) resummation \cite{Braaten:1989mz,Manuel:1995td,Blaizot:2001nr}.
The second type, in contrast, refers to the screening of static temporal gluonic modes $A_{\nu=0}$, which is minimally described by the dimensionally reduced effective theory known as Electrostatic QCD (EQCD) \cite{Braaten:1995cm,Braaten:1995jr,Appelquist:1981vg,Kajantie:1995dw}.
While the first type of screening persists even at zero temperature, the latter arises exclusively at nonzero temperature. Their relative importance in observables such as the thermodynamic pressure can be estimated by the ratio $T/m_\text{E}$ \cite{Kurkela:2016was}. 

Since the combination of diagrams in Eq.~\eqref{eq:bugblatterdiags} has been argued to be IR finite in the $T = 0$ limit based on the properties of the HTL three-gluon interaction \cite{Moore:2023glb} (and shown explicitly in Ref.~\cite{Navarrete:2024zgz}), the nonstatic gluonic contributions are free of divergences at $O(\alpha_s^3)$ even at nonzero temperatures. However, the lower dimensionality of the static sector at $T > 0$---as described by the EQCD effective theory---leads to an enhanced IR sensitivity that translates into genuine divergences.

Following previous work in thermal field theory \cite{Kurkela:2016was,Navarrete:2025yxy}, a simple handling of soft modes for $\Delta p$ can be established through the expression
\begin{equation}
\label{eq:resummedp}
\Delta p = (\Delta p^{\text{naive}} - \Delta p^{\text{naive}}_{\text{EQCD}}) + \Delta p^{\text{res}}_{\text{EQCD}}.
\end{equation}
Here, the $\Delta p^\text{naive}$ contribution to $O(\alpha_s^3)$ corresponds to the Bugblatter diagrams displayed in Eq.~\eqref{eq:bugblatterdiags}, while the $\Delta p_\text{EQCD}^\text{res}$ term captures the physics of the soft scale in a fully resummed, infrared-finite fashion in EQCD. Finally, the $\Delta p_\text{EQCD}^\text{naive}$ quantity is obtained by expanding resummed contributions in $\Delta p_\text{EQCD}^\text{res}$ in powers of $\alpha_s$, resulting in integrals that act as local counterterms that eliminate any IR-sensitive contributions arising in the loop expansion $\Delta p^\text{naive}$. A more detailed explanation for each of these terms is provided in Appendix~\ref{appsec:EQCD}.

Due to the multiscale nature of the problem involving $\muq$ and the ratio $m_\text{E}/\muq \propto \alpha_s^{1/2}$, the weak-coupling expansion of $\Delta p$ exhibits nonanalytic dependence in $\alpha_s$---particularly logarithms and fractional powers. 
In this work, we achieve the evaluation of $\Delta p$ in Eq.~\eqref{eq:resummedp} to $O(\alpha_s^{7/2})$ accuracy in the renormalized coupling, adopting the form 
\begin{equation}
\label{eq:resummedpresult}
\Delta p = \left(\frac{\alpha_s}{\pi}\right)^3\left[ c_1 + c_2 \ln \frac{\alpha_s}{\pi} + c_3 \left(\frac{\alpha_s}{\pi}\right)^{1/2} \right],
\end{equation}
where the coefficients $c_i=c_i(T, \muq)$ are finite functions. The coefficients $c_2$ and $c_3$ arise from the EQCD terms in Eq.~\eqref{eq:resummedp} and are computable analytically as simple functions of $T$ and $\muq$. However, $c_1$ must exhibit a nontrivial functional dependence as it arises from the integral expression $\Delta p^\text{naive}$ in full QCD. Note that $\Delta p \to 0$ as $\muq \to 0$, since it arises explicitly as a difference between QCD and the PQ theory with a $\muq \mapsto -\muq$ symmetry.

The precise definitions for the $c_i$ coefficients and their evaluation can be found in greater detail in Appendix~\ref{appsec:EQCD}; here we content ourselves to list the results and describe what the computation entails. In EQCD, the presence of a nonzero quark chemical potential breaks the charge conjugation symmetry, acting as $\bar{A}_0\mapsto-\bar{A}_0$ in the low-energy temporal gauge fields $\bar{A}_0$.
This gives rise to a renormalizable effective operator $i\bar{\gamma} \,\text{tr}\,(\bar{A}_0)^3$ in EQCD \cite{Hart:2000ha}, where the coefficient $\bar{\gamma}$ is derived from the matching relation in Eq.~\eqref{eq:gammamatching}.
Crucially, this operator is absent in the PQ version of EQCD, as such a theory exhibits charge conjugation symmetry by construction \footnote{This accords with the fact that, by symmetry, the color trace in this operator is proportional to the $d_{abc}$ structure.}. Hence, soft contributions to $\Delta p$ up to $O(\alpha_s^{7/2})$ are graphs with two insertions of cubic $\bar{A}_0$ vertices, corresponding to two- and three-loop vacuum diagrams in EQCD. 

The leading-order contribution to $\Delta p$ in EQCD is the fully massive two-loop sunset diagram in Eq.~\eqref{eq:resleading}, which features a logarithmic divergence that is responsible for the $\ln \alpha_s$ structure and hence the $c_2$ coefficient in Eq.~\eqref{eq:resummedpresult}. On the other hand, the coefficient $c_3$ arises from the evaluation of the four three-loop graphs in EQCD listed in Eq.~\eqref{eq:g7res}, of which only two are nontrivial. The resulting analytical expressions for $c_2$ and $c_3$ for general $N_c$ and $N_f$ can be found in Eqs.~\eqref{eq:logalpha} and \eqref{eq:as7/2}, respectively.
Here we set $N_c=N_f=3$, which yields
\begin{equation}
\begin{split}
\label{eq:c2c3}
    c_2 = -5 T^2\muq^2, \quad  c_3 = 48.8753 \frac{T^2\muq^2}{\sqrt{1+\left(\frac{\mu_\text{q}}{\pi T}\right)^2}},
\end{split}
\end{equation}

Finally, the coefficient $c_1$ includes contributions from EQCD and full QCD. 
The former arises from the counterterm $\Delta p^\text{naive}_\text{EQCD}$, whose explicit form is shown in Eq.~\eqref{eq:naiveEQCD}. This counterterm cancels an IR sensitivity from the diagrams in Eq.~\eqref{eq:bugblatterdiags} on the integrand level. Together with the built-in local cancellation of UV divergences in the sum of diagrams in Eq.~\eqref{eq:bugblatterdiags}, this leads to a four-loop integration that is locally finite in all kinematic regions. 

Although the full diagrammatic expression is locally fully finite after including the proper IR counterterm from EQCD, its complexity is apparent.
Evaluating the corresponding scalarized sum-integrals that emerge by expanding the Dirac traces requires knowledge of several nonfactorizable four-loop vacuum integral topologies at $T, \mub > 0$ in QCD \cite{Navarrete:2024zgz}, computations that have thus far remained intractable.
To bypass this technical obstacle altogether, we employ at the full-diagram level an algorithmic approach originally developed in collider physics based on the Loop-Tree Duality (LTD), whose main goal is to render loop integrals numerically tractable through direct Monte Carlo (MC) integration in momentum space (see, e.g., Refs.~\cite{Soper:1999xk,Catani:2008xa,Becker:2012bi,Buchta:2015wna,Runkel:2019yrs,Capatti:2019ypt,Capatti:2019edf,Driencourt-Mangin:2019aix,Aguilera-Verdugo:2020set,Capatti:2020ytd,Capatti:2020xjc,Capatti:2022tit,Ramirez-Uribe:2024rjg,LTD:2024yrb}).
 
\begin{table}[t]
\centering
\begin{tabular}{cccc}
    \toprule
   $c_1(T,\mu_\text{q})$ [GeV$^4$] & $T$ [MeV] & $N\,[10^8]$ & $[\mathrm{ms}]$ \\ 
    \midrule
     0.2401(8) & 25 & 4.62 & 7.2 \\
     0.1853(7) & 50 & 2.9 & 7.4 \\
     -0.09523(31) & 125 & 2.9 & 7.6 \\
     -0.19729(20) & 175 & 2.9 & 9.5 \\
    \bottomrule
\end{tabular}
\caption{Representative values for $c_1(T,\mu_\text{q})$ along the arc defined by $(4.338\pi T)^2+(3\mu_\text{q})^2=(3\text{ GeV})^2$. Here, $N$ indicates the number of MC samples needed for each $(T,\mu_\text{q})$ point with $[\mathrm{m s}]$ the time spent on one sample on a single CPU core.}
\label{table:c1}
\end{table}

The LTD approach was recently generalized to a thermal setting---first at nonzero density \cite{Navarrete:2024zgz} and later at finite temperature \cite{Navarrete:2025yxy}---and has proven effective already at the four-loop level \cite{Karkkainen:2025nkz}.
While $c_1$ in Eq.~\eqref{eq:resummedpresult} was computed with this method already in Ref.~\cite{Navarrete:2024zgz} at $T=0$, here we achieve its evaluation using a full extension to nonzero $T$, dubbed \texttt{thermalLTD} \cite{Capatti}.
The general algorithm consists roughly of three steps \cite{Navarrete:2025mnt, KSPhD}:
(i) a subtraction of IR and UV divergences at the integrand level;
(ii) an analytic evaluation of Matsubara sums, yielding a closed-form locally finite three-dimensional representation; and
(iii) a Monte Carlo integration of the resulting phase-space integral using \texttt{Symbolica} \cite{symbolica}.
In Table~\ref{table:c1}, some representative numerical values for $c_1$ along an arc in the phase diagram are shown, together with the total number of MC samples and the evaluation time per sample per core using \texttt{thermalLTD}. The total CPU time for each point is obtained by multiplying these two quantities.

\section{Results and Discussion}

\begin{figure}
    \centering \includegraphics[width=1.05\linewidth]{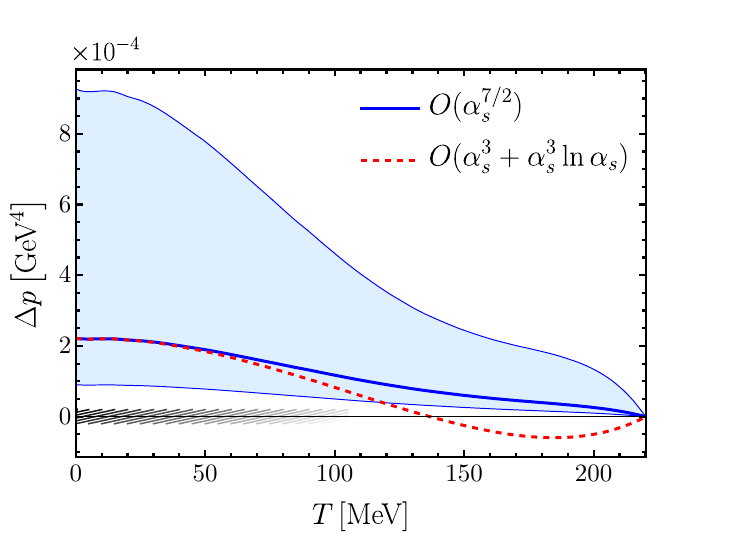}
    \caption{Pressure difference $\Delta p$ for $N_f = 3$ from Eq.~\eqref{eq:resummedpresult} throughout the phase-diagram arc $(4.338\pi T)^2+(3\muq)^2=(3\text{ GeV})^2$ for $T\in [0,220]$ MeV. The renormalization scale is chosen as $\bar{\Lambda} = X \sqrt{(2 \muq)^2 + (0.723 \times 4 \pi T)^2}$, where $X$ is varied in the range $[1/2,2]$ for the full $\Delta p$ and fixed to $X=1$ for the partial result with $c_3=0$. The numerical errors from the MC integration for $c_1$ are very small and hence not shown here. The hatching on the $T$ axis for $T\lesssim T_c^\text{PQ}$ refers to the range where potentially large pairing effects might occur.}
     \label{fig:fullpressure}
\end{figure}

Collecting our results for the $c_i$ coefficients appearing in Eq.~\eqref{eq:resummedpresult}, in Fig.~\ref{fig:fullpressure} we show the $N_f = 3$ $\Delta p$ along the arc ${(4.338\pi T)^2+(3\mu_\text{q})^2=(3\text{ GeV})^2}$ in the phase diagram for the temperature range ${T\in[0,220] \text{ MeV}}$, with and without the $c_3$ term. Given our choice of renormalization scale, this arc is such that $\alpha_s$ remains constant and relatively small ($\alpha_s \lesssim 0.48$), while covering points inside the region of validity of current lattice simulations in full QCD \cite{HotQCD:2018pds,Borsanyi:2020fev,Abuali:2025tbd} (see Fig.~\ref{fig:QCDpD}). We stress that $\Delta p$, while well-defined for $T\lesssim T_c^\text{PQ}$ (hatched interval in Fig.~\ref{fig:fullpressure}), may receive sizable pairing contributions from the phase-quenched sector.

In Fig.~\ref{fig:fullpressure}, we see that already for $T\gtrsim 140 \text{ MeV}$, corresponding to $\mu_\text{q}\lesssim 770 \text{ MeV}$, the 3-loop EQCD contribution is significant, ultimately preventing $\Delta p$ from crossing into a region where the nonperturbative inequality $\Delta p\geq 0$ is violated \footnote{In the $\alpha_s\to0$ limit, the $O(\alpha_s^3\ln \alpha_s)$ term is positive and dominant for all $T$ and $\mu_\text{q}$. 
Therefore, $\Delta p$ approaches zero from above in the weak-coupling regime.}. 
In this regime, the typical momenta of nonstatic (hard) and static (soft) thermal gluons are of similar magnitude ($2\pi T \approx m_\text{E}$), giving rise to large soft corrections. 
This showcases the importance of including soft contributions in the form of fractional powers in $\alpha_s$ in high-temperature observables \cite{Braaten:1995ju,Blaizot:2003iq}.

Upon inspecting the scale uncertainties in Fig.~\ref{fig:fullpressure} for the complete $O(\alpha_s^{7/2})$ result, we find that the band decreases as one moves away from the $T=0$ axis and effectively vanishes near $T\simeq 220\,\text{MeV}$. This reflects the disappearance of $\Delta p$ in the limit $\mu_\text{q} \to 0$. The small-$\mu_\text{q}$ region therefore provides controlled benchmark points for direct comparison between PQ lattice simulations and full QCD results valid for $\mu_\text{B} \lesssim 3.5T$ \cite{Abuali:2025tbd}. Given the role of the $c_3$ term at higher temperatures, a quantitative assessment of uncertainties across the phase diagram ultimately requires the determination of $O(\alpha_s^4)$ corrections to $\Delta p$.

Having motivated the need to extend the perturbative calculation to the next order, we emphasize that this task is conceptually straightforward. Importantly, extending our calculation to $O(\alpha_s^{4})$ is significantly simpler than computing the full QCD pressure to the same order: working with $\Delta p$ reduces the number of required 5-loop diagrams by roughly an order of magnitude, which can then be evaluated with \texttt{thermalLTD} given enough computational resources.
Furthermore, the soft contributions at $O(\alpha_s^{4}+\alpha_s^{4}\ln\alpha_s)$ and $O(\alpha_s^{9/2})$ can be computed within the effective theories with significantly less effort. Notably, the onset of nonperturbative effects in $\Delta p$ is postponed to $O(\alpha_s^5)$, as purely gluonic diagrams sensitive to the magnetic scale $\alpha_s T$ \cite{Linde:1980ts, DiRenzo:2006nh} cancel. A more detailed discussion is given in Appendix~\ref{appsec:deltaphighorder}.

As a further generalization of our computation of $\Delta p$, a nonzero strange-quark mass can be readily incorporated into \texttt{thermalLTD} and the EQCD theory. We note, however, that such effects on the pressure have previously been found to be of $O(1\%)$ in the hot and dense limits of QCD \cite{Laine:2006cp,Kurkela:2009gj,Gorda:2021gha}. Whether this conclusion also applies to $\Delta p$ in the unpaired phase remains an interesting question for future work.

Finally, we stress that the values of $(T,\mu_\text{q})$ shown in Fig.~\ref{fig:fullpressure} were chosen for illustration. Indeed, one can trivially compute the value of $\Delta p$ at another point in the phase diagram from these values, provided $\alpha_s$ remains small, by noting that the coefficients $c_i$ scale homogeneously as $c_i(s T, s \muq) = s^4 c_i(T, \muq)$.
Because pairing effects are suppressed for $T\gtrsim T_c^{\mathrm{PQ}}$, our results demonstrate that a large region of the QCD phase diagram can be determined to high accuracy once $p_{\mathrm{PQ}}$ is known. For this reason, we strongly encourage the lattice community to perform this computation.

\begin{acknowledgments}
We thank Gergely Endrődi, Valentin Hirschi, Guy Moore, Marco Panero, Tobias Rindlisbacher, Kari Rummukainen, York Schröder, and Aleksi Vuorinen for helpful comments and suggestions. T.G. was supported in part by the Deutsche Forschungsgemeinschaft (DFG, German Research Foundation) through the CRC-TR 211 `Strong-interaction matter under extreme conditions'-- project number 315477589 -- TRR 211 and by the ERC Advanced Grant ``JETSET: Launching, propagation and emission of relativistic jets from binary mergers and across mass scales'' (Grant No.~884631). 
P.N., R.P., L.S. and K.S.
have been supported by the Research Council of Finland grants 347499,
353772, 354533, and 354572, and by the ERC Consolidator Grant ExPertQCD (grant No.~101231521).  K.S. also gratefully acknowledges support from the Albert Einstein Center for Fundamental Physics (AEC) at the University of Bern. Computational resources have been
provided by CSC - IT Center for Science, Finland.
\end{acknowledgments}

\appendix

\vspace{0.5cm}

\section{Phase-quenched pairing gap}
\label{appsec:PQpairing}

In this Appendix, we provide details on the derivation of the quark pairing gap and the critical temperature in the PQ theory for general $N_f$.

Introducing the gap structure into the half-flavor inverse propagator matrix leads to the ansatz
\begin{equation}
\label{eq:prop_matrix}
\begin{pmatrix}
    \langle q_+ \bar{q}_+ \rangle & \langle q_+ \bar{q}_- \rangle \\
     \langle q_- \bar{q}_+ \rangle & \langle q_- \bar{q}_- \rangle 
\end{pmatrix}^{-1}
= \ \ 
\begin{pmatrix}
[G^+]^{-1}\mathds{1}_\mathrm{f}&+\Phi^\dagger \gamma^5 \mathds{1}_\mathrm{c}\\
-\Phi\gamma^5 \mathds{1}_\mathrm{c} &[G^-]^{-1}\mathds{1}_\mathrm{f}
\end{pmatrix},
\end{equation}
where $\mathds{1}_\mathrm{f}$  and $\mathds{1}_\mathrm{c}$ are identity matrices in flavor and color space, respectively, and $G^\pm$ is the half-fermion propagator of $q_\pm$ in the case of a vanishing gap. Further, we can assume the flavor angles $\Phi=\Delta^\text{PQ}\mathds{1}_\mathrm{f}$ to be diagonal.
This can always be achieved by performing a singular-value decomposition of the matrix $\Phi$, which amounts to performing two unitary flavor transformations $q_+\mapsto U_+ q_+$ and $q_- \mapsto U_- q_-$ to diagonalize $\Phi$. 
Since there is no difference between the fermions, the singular values of $\Phi$ must be equal, and we find $N_f$ identical gap equations corresponding to $N_f$ identical pairings $\langle q_{-}^f \gamma^5 \mathds{1}_c \bar{q}_{+}^f \rangle$ between a single $q^f_+$ of flavor $f$ with a single $q^f_-$ of the same flavor.

To set up the gap equation, we require the free propagator $S_0$ and the gluon interaction matrix $\Gamma^\mu_a$ in our chosen basis. These are
\begin{equation}\label{eq:freeprop}
\begin{split}
S_0^{-1}\, &\coloneq \, 
\begin{pmatrix}
    [G^+_0]^{-1}&0\\
0&[G^-_0]^{-1}
\end{pmatrix}\otimes \mathds{1}_{f},
 \\ 
    \Gamma^\mu_a\, &\coloneq \, 
\begin{pmatrix}
    \gamma^\mu T_a&0\\
    0& \gamma^\mu T_a\\
\end{pmatrix}\otimes \mathds{1}_{f},
\end{split}
\end{equation}
where
\begin{equation}
[G^\pm_0]^{-1}=\underset{e=\pm}{\sum}[k_0\pm(\muq-ek)]\gamma^0\Lambda^{(\pm e)}_{\vec{k}}
\end{equation}
is the massless Dirac operator, and the projection ${\Lambda^{(e)}_{\vec{k}} = (1+e\gamma^0\vec{\gamma}\cdot\hat{k})/2}$ provides the spinor structure.
Furthermore, the full propagator $S$ and self-energy contributions $\Sigma$ are denoted by
\begin{equation} \label{eq:fullprop}
    S \coloneq
\begin{pmatrix}
    G^+&F^+\\
F^-&G^-
\end{pmatrix}\otimes \mathds{1}_{f},
\ \quad 
\Sigma  \coloneq  
\begin{pmatrix}
    \Sigma^+&\Phi^+\\
\Phi^-&\Sigma^-
\end{pmatrix}\otimes \mathds{1}_{f},
\end{equation}
where we can factor out the flavor structure due to the diagonal form of $\Phi$. The off-diagonal elements $F^\pm$ and $\Phi^\pm$ are related to the off-diagonal elements of Eq.~(\ref{eq:prop_matrix}), and will be fixed later. 

We derive the self-consistent Schwinger-Dyson equation following the effective action approach of Ref.~\cite{Cornwall:1974vz}, with a slight modification regarding the PQ half-flavors. The half-flavor contribution to the two-particle-irreducible (2PI) PQ effective action is stated as
\begin{equation}
    \Gamma[S] \coloneq \frac{1}{2}\mathrm{Tr}[\ln S^{-1}] +\frac{1}{2}\mathrm{Tr}[S_0^{-1}S -1] +\Gamma_2[S,D],
\end{equation}
where $\Gamma_2[S,D]$ are 2PI diagrams constructed by the half-flavor propagator $S$ and gluon propagator $D$ with additional factors of $1/2$ introduced for each fermion loop. The stationary point of the effective action with respect to the half-flavor propagator $S$ leads to the following equations:
\begin{equation}\label{eq:SDeq}
    \frac{1}{2}S^{-1} = \frac{1}{2} S_0^{-1} + \frac{1}{2}\Sigma,\quad 
    \frac{1}{2}\Sigma = \frac{\delta \Gamma_2}{\delta  S}[S,D],
\end{equation}
which serve as the self-consistency equations. 
At leading order, $\Gamma_2[S,D]$ is simply given by the two-loop sunset diagram---taking its variational derivative results in the self-energy
\begin{equation}\label{eq:selfenergy}
 \frac{1}{2}\Sigma(K) = -\frac{g^2}{2}\SumInt_{Q} \Gamma_a^\mu S(Q) \Gamma_b^\nu  D_{\mu\nu}^{ab}(K-Q).
\end{equation}
The diagonal self-energy components
\begin{equation}
\Sigma^\pm(K) \approx \bar{g}^2\log( {M^2}/{k_0^2} ) k_0\gamma^0 \Lambda_{\vec{k}}^\pm
\end{equation}
were already given in Ref.~\cite{Fujimoto:2023mvc}, with $M^2 \equiv N_f g^2\muq^2/(8\pi)$ and $\bar{g}\equiv g/(3\sqrt{2}\pi)$. Using the $\Phi^+$ component of the self-consistency equation, it is possible to eliminate the dependence on $G^\pm$ and $F^\pm$ via the other components of the same equation, thus arriving at an expression given in terms of $\Phi^\pm$ and the gluon propagator $D$. Here, our structure of the pairing gap $\Phi^\pm_k = \pm \Delta_k \gamma^5 \mathds{1}_c$, which is fixed by Eq.~(\ref{eq:prop_matrix}), can be used to solve for $\Delta_k$.

Carrying out a HTL approximation for the gluon propagator $D$ along the lines of Ref.~\cite{Pisarski:1999tv}, the Matsubara sum in the self-energy is left to be solved. We achieve this by splitting up the propagator into longitudinal and transverse components, and making use of the formulas for the respective Matsubara sum contribution derived in Ref.~\cite{Pisarski:1999tv}, which hold in the limit $g\mu \gg \Delta_k$.

With all these ingredients, we arrive at the following gap equation:
\begin{equation}
\begin{split}
\label{eq:gapeq}
    \Delta_k =
    2\bar{g}^2\int_0^\delta \frac{\mathrm{d}(p-\mu)}{\tilde{\epsilon}_p}\Bigg[& Z^2(\tilde{\epsilon}_p) \tanh\Bigg(\frac{\tilde{\epsilon}_p}{2T}\Bigg) \\
    &\times\frac{1}{2}  \ln\Bigg( \frac{\tilde{b}^2\muq^2}{|\tilde{\epsilon}_p^2-\tilde{\epsilon}_k^2|}\Bigg)\Delta_p
    \Bigg],
\end{split}
\end{equation}
where $\tilde{b}=256\pi^4g^{-5}(2/N_f)^{\frac{5}{2}}$, the wavefunction renormalization factor is $Z(p_0) = [1+\bar{g}^2\ln(M^2/p_0^2)]^{-1}$, and the dispersion relation reads $\epsilon_p = \sqrt{(p-\muq)^2+\Delta_p^2}$; the modified dispersion relations are $\tilde{\epsilon}_p = \epsilon_p Z(\epsilon_p)$, while $\delta$ gives an arbitrary energy scale $\Delta_k\ll\delta\ll g\muq$ encoding the momenta around the Fermi surface. 

In Ref.~\cite{Fujimoto:2023mvc}, Eq.~\eqref{eq:gapeq} was already considered in the $N_f=2$ case. Here, we are able to reproduce the result in the $N_f=2$ case and generalize the pairing gap estimate $\Delta^{\text{PQ}}$ to arbitrary number of flavors $N_f$. Using the gap equation and the methods used in Ref.~\cite{Pisarski:1999tv}, we obtain the PQ critical temperature reported in the main text.

\section{EQCD contributions to $\Delta p$}
\label{appsec:EQCD}

The purpose of this Appendix is threefold: (i) compute the $O(\alpha_s^3\ln\alpha_s)$ contribution to $\Delta p$ and derive $\Delta p^\text{naive}_\text{EQCD}$ thereof; (ii) construct the locally finite $O(\alpha_s^3)$ expression to be integrated with \texttt{thermalLTD}; and (iii) evaluate the $O(\alpha_s^{7/2})$ contribution to $\Delta p$. This fixes the coefficients $c_1$, $c_2$ and $c_3$ appearing in Eq.~\eqref{eq:resummedpresult}.

To this order, the effective theory we need is \cite{Hart:2000ha}
\begin{equation}
\label{appeq:EQCDlagrangian}
\begin{split}
    L_\text{EQCD}&=\frac{1}{2}\text{tr}\,\bar{F}_{ij} \bar{F}_{ij} + \text{tr}\,[\bar{D}_i,\bar{A}_0][\bar{D}_i,\bar{A}_0] \\ 
    &+ m_\text{E}^2\,\text{tr}\,\bar{A}_0^2+ i\bar{\gamma} \,\text{tr}\,\bar{A}_0^3,
\end{split}
\end{equation}
where the field-strength tensor is $\bar{F}_{ij}\equiv \partial_i \bar{A}_j-\partial_j \bar{A}_i-ig_\text{E}[\bar{A}_i,\bar{A}_j]$ and the covariant derivative in the fundamental representation is $\bar{D}_i\equiv \partial_i-ig_\text{E}\bar{A}_i$. The barred low-energy fields have mass dimension one-half, the dimensionful coupling is $g^2_\text{E}=g^2T$ to leading order, and the mass parameter to leading order is that of Eq.~\eqref{eq:screeningmass}.

To use $\Delta p^\text{naive}_\text{EQCD}$ as an infrared counterterm in the evaluation of $c_1$ in Eq.~\eqref{eq:resummedpresult}, the $\bar{\gamma}$ effective parameter is required in unintegrated form. To this end, we match static three-point functions for the $A_0$ field in QCD and EQCD to one-loop order at vanishing external momenta $P_i=0$, where the Euclidean four-momenta is $P =(\omega^\text{b/f}_n,\mathbf{p})$ for bosons ($\omega_n^\text{b}=2\pi Tn$) and fermions ($\omega_n^\text{f}=\pi T(2n+1)+i\mu_\text{q}$). This results in the following matching equation:
\begin{widetext}
\begin{equation}
\label{eq:gammamatching}
    -3!i\overline{\gamma}d^{abc} = 0 + \raisebox{-0.65cm}{\includegraphics[scale=0.9]{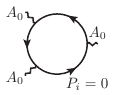}} + \raisebox{-0.65cm}{\includegraphics[scale=0.9]{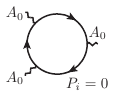}} + \text{(gluon/ghost loops)} + O(g^5),
\end{equation}
\end{widetext}
where the purely bosonic loops vanish. This yields
\begin{equation}
    \bar{\gamma}=\frac{i}{3}g^3N_fT^{1/2} \Gamma^{(3)}_{000}(0,0)+O(g^5),
\end{equation}
where the one-loop sum-integral is defined as
\begin{equation}
    \Gamma^{(3)}_{\mu\nu\sigma}(P_1,P_2) \coloneq \SumInt_{\{S\}}\frac{\text{tr}\,[\slashed{S}\gamma_\mu(\slashed{S}-\slashed{P}_1)\gamma_{\nu}(\slashed{S}+\slashed{P}_2)\gamma_\sigma]}{S^2(S-P_1)^2(S+P_2)^2},
\end{equation}
which for vanishing external momenta reduces to
\begin{equation}
\begin{split}
\label{eq:gamma3}
    \Gamma^{(3)}_{\mu\nu\sigma}(0,0) &= 4(1-d)U_\mu U_\nu U_\sigma\SumInt_{\{S\}}\frac{S_0}{(S^2)^2}.
\end{split}
\end{equation}
Here, $U_\mu\equiv(1,\mathbf{0})$ is the four-vector defining the rest frame of the heat bath, and the sum-integral symbol is defined as follows:
\begin{equation}
\begin{split}
    \SumInt_{P/\{P\}}f(P) &\coloneq T\sum_{n \in \mathbb{Z}}\int_\mathbf{p} f(\omega^\text{b/f}_n, \mathbf{p}), \\
    \int_\mathbf{p} &\coloneq \left( \frac{e^{\gamma_\text{E}}\bar{\Lambda}}{4\pi} \right)^{\varepsilon} \int \frac{\mathrm{d}^d\mathbf{p}}{(2\pi)^d},
\end{split}
\end{equation}
where we employ dimensional regularization in $d=3-2\varepsilon$ spatial dimensions in the $\overline{\text{MS}}$ renormalization scheme. The sum-integral in Eq.~\eqref{eq:gamma3} is straightforward to evaluate for general $d$ in terms of Hurwitz Zeta functions (see, e.g., Ref.~\cite{Vuorinen:2003fs}), resulting in the effective parameter $\bar{\gamma}=g^3N_fT^{1/2}\mu_\text{q}/3\pi^2+O(\varepsilon)$ \cite{Hart:2000ha}. Although the linear term in $\varepsilon$ may result in $O(1)$ contributions in individual terms by multiplying a pole in $\varepsilon$, it can be seen to drop out in the physical result $\Delta p$. To this end, we keep $\bar{\gamma}$ in $d$-dimensional form in intermediate steps.

The leading-order soft contribution to $\Delta p$ is the two-loop sunset diagram in EQCD, which reads
\begin{equation}
\begin{split}
\label{eq:resleading}
    \Delta &p^\text{res}_\text{EQCD} = -\raisebox{-0.42cm}{\includegraphics[scale=0.8]{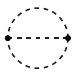}} + O(g^7) \\
    &= \frac{3}{16}\bar{\gamma}^2T d^{abc}d^{abc}S_\text{3d}(m_\text{E}) + O(g^7),
\end{split}
\end{equation}
where the minus sign in front comes from the definition of $\Delta p$, and the corresponding massive three-dimensional two-loop integral is ($\mathbf{r}=\mathbf{p}-\mathbf{q}$) \cite{Farakos:1994kx}
\begin{equation}
\begin{split}
    S_\text{3d}(m) &\coloneq \int_{\mathbf{pq}}\frac{1}{(\mathbf{p}^2+m^2)(\mathbf{q}^2+m^2)(\mathbf{r}^2+m^2)} \\
    &= \frac{1}{(4\pi)^2}\left( \frac{1}{4\varepsilon}+\ln\frac{\bar{\Lambda}_\text{3d}}{3m}+\frac{1}{2}+O(\varepsilon) \right).
\end{split}
\end{equation}
Here, the $\overline{\text{MS}}$ scale is that of the three-dimensional theory. 
Before commenting on the resulting expression, let us derive the $\Delta p^\text{naive}_\text{EQCD}$ contribution from $\Delta p^\text{res}_\text{EQCD}$.

In Eq.~\eqref{eq:resleading}, we need to expand $m^2_\text{E}=0+O(g^2)$ in the propagators and write $\bar{\gamma}$ in unintegrated form. At the order we are working, it suffices to simply set $m_\text{E}=0$, yielding the expression
\begin{equation}
\label{eq:naiveEQCD}
    \Delta p^\text{naive}_\text{EQCD} = \frac{g^6 T^2}{3\times 4^2}d^{abc}d^{abc}\left( \Gamma^{(3)}_{000} \right)^2S_\text{3d}(0) + O(g^8).
\end{equation}
Here, the $S_\text{3d}(0)$ integral is logarithmically divergent in the infrared, counteracting the infrared sensitivity exhibited by the Bugblatter diagrams in $\Delta p^\text{naive}$. We stress that, in order to remove this sensitivity at the local (i.e., integrand) level, one must rely on the full expressions for $\Gamma^{(3)}_{000}$ and $S_\text{3d}$ without any further manipulations.

From Refs.~\cite{Navarrete:2024zgz,Moore:2023glb}, it is straightforward to write down the finite-temperature generalization of the Bugblatter diagrams defining the $\Delta p^\text{naive}$ quantity. It reads
\begin{equation}
    \Delta p^\text{naive} = \frac{g^6}{48}N_f^2d^{abc}d^{abc}\SumInt_{P_1P_2}\frac{[\text{Im}\,\Gamma^{(3)}_{\mu\nu\sigma}(P_1,P_2)]^2}{P_1^2 P_2^2 (P_1-P_2)^2}.
    \label{eq:fullnaive}
\end{equation}
Importantly, despite $\Delta p^\text{naive}$ being an ultraviolet-finite quantity (as explained in Refs.~\cite{Navarrete:2024zgz,Moore:2023glb}; the argument remains valid at nonzero temperature), the infrared counterterm $\Delta p^\text{naive}_\text{EQCD}$ itself suffers from ultraviolet divergences. Therefore, an adequate counterterm $\Delta p^\text{naive}_\text{EQCD,CT}$ must be further subtracted at the integrand level. Although straightforward to derive in the present case, such subtractions can be constructed algorithmically e.g. via the recursive $R$-operation (see, e.g., Ref.~\cite{KSPhD}). This leads us to rewrite $\Delta p$ in Eq.~\eqref{eq:resummedp} in the form
\begin{equation}
\label{eq:resummedp2}
\begin{split}
    \Delta p &= \left( \Delta p^{\text{naive}} - \Delta p^{\text{naive}}_{\text{EQCD}} + \Delta p^{\text{naive}}_{\text{EQCD,CT}} \right) \\
    &+ \left(\Delta p^{\text{res}}_{\text{EQCD}} - \Delta p^{\text{naive}}_{\text{EQCD,CT}} \right).
\end{split}
\end{equation}
With all ingredients in place, the combination in the first parentheses of Eq.~\eqref{eq:resummedp2} is in a form that can be integrated with \texttt{thermalLTD}, providing us with the coefficient $c_1$ in the tabulated form displayed in Table~\ref{table:c1}.

The counterterm $\Delta p^\text{naive}_\text{EQCD,CT}$ in the last term of Eq.~\eqref{eq:resummedp2} must be evaluated in the $\overline{\text{MS}}$ scheme. Our choice of subtraction leads to the result
\begin{equation}
\begin{split}
\label{eq:naiveEQCDCT}
    \Delta p^\text{naive}_\text{EQCD,CT} &= \frac{3}{16}\bar{\gamma}^2T d^{abc}d^{abc}\frac{1}{(4\pi)^2} \\
    &\times\left( \frac{1}{4\varepsilon}+\ln\frac{\bar{\Lambda}_\text{3d}}{3m_\text{UV}}+\frac{1}{2}+O(\varepsilon) \right),
\end{split}
\end{equation}
where $m_\text{UV}=2 \pi T$ is a convenient infrared regulator that drops out in $\Delta p$. This expression can be combined with Eq.~\eqref{eq:resleading} as in the last term of Eq.~\eqref{eq:resummedp2}, resulting in
\begin{equation}
\begin{split}
\label{eq:logalpha}
    \Delta p^{\text{res}}_{\text{EQCD}} - \Delta p^{\text{naive}}_{\text{EQCD,CT}} &= \frac{\mu_\text{q}^2 T^2}{12}\left( \frac{\alpha_s}{\pi} \right)^3  N_f^2d^{abc}d^{abc} \\
    &\times \ln \frac{m_\text{UV}}{m_\text{E}} + O(\alpha_s^{7/2}).
\end{split}
\end{equation}
Here, we witness the cancellation of poles in $\varepsilon$ and the $\bar{\Lambda}_\text{3d}$ scale, leaving a finite $\ln m_\text{E}/m_\text{UV}\sim \ln\alpha_s$ term.
This provides the coefficient $c_2$ reported in Eq.~\eqref{eq:c2c3}, in addition to a finite contribution to $c_1$.

Finally, we discuss the evaluation of the coefficient $c_3$. The relevant three-loop diagrams in EQCD consist of two insertions of cubic $\bar{A}_0$ vertices and one spatial $\bar{A}_i$ line, leading to the following diagrammatic expansion:
\begin{equation}
\label{eq:g7res}
    -\Delta p^\text{res}_\text{EQCD}\biggr\rvert_{\alpha_s^{7/2}}=    \raisebox{-0.42cm}{\includegraphics[scale=0.8]{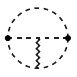}} + \raisebox{-0.42cm}{\includegraphics[scale=0.8]{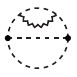}} + \raisebox{-0.42cm}{\includegraphics[scale=0.8]{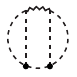}} + \raisebox{-0.42cm}{\includegraphics[scale=0.8]{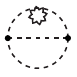}},
\end{equation}
where the wavy lines represent $\bar{A}_i$ fields. The third and fourth diagrams can immediately seen to vanish---the former due to a color contraction of the type $d_{abc}f_{abe}=0$ (here, $f_{abc}=-2i\,\text{tr}([T_a,T_b],T_c)$ is the totally antisymmetric $\text{SU}(N_c)$ structure constant), and the latter due to a factorization onto a scale-free $\bar{A}_i$ tadpole that is set to zero in dimensional regularization. Therefore, only two diagrams are nontrivial, and can be seen to be proportional to the color structure $N_c d_{abc}d_{abc}$.

We carry out the computation of Eq.~\eqref{eq:g7res} in a general covariant gauge parametrized by $\xi$, entering the $\bar{A}_i$ propagator in the form $\sim (\delta_{ij}-\xi p_i p_j/\mathbf{p}^2)$. Given that $m_\text{E}$ is the sole scale in loop integrals, we can trivially factor it out. With the help of the computer algebra package \texttt{FORM} \cite{Vermaseren:2000nd} and symmetrizations along the lines of Ref.~\cite{Navarrete:2024ruu}, the nontrivial diagrams yield the following \footnote{As a further check, we have generated \textit{all} three-loop graphs in EQCD at super-renormalizable order and isolated those contributing to the $d^{abc}d^{abc}$ group-theory structure, leading exactly to Eq.~\eqref{eq:diagsEQCDg7}}:
\begin{widetext}
\begin{equation}
\begin{split}
    \raisebox{-0.55cm}{\includegraphics[scale=1]{figs/EQCD_g7_1.pdf}} &= \frac{9}{32} \overline{\gamma}^2 \frac{g_\text{E}^2}{m_\text{E}} N_c d_{abc} d_{abc} \left( -4\raisebox{-0.55cm}{\includegraphics[scale=1]{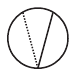}}+2\raisebox{-0.55cm}{\includegraphics[scale=1]{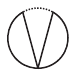}}+\raisebox{-0.55cm}{\includegraphics[scale=1]{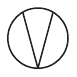}}+2\raisebox{-0.55cm}{\includegraphics[scale=1]{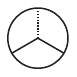}}-2\xi \raisebox{-0.55cm}{\includegraphics[scale=1]{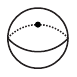}} \right), \\
    \raisebox{-0.55cm}{\includegraphics[scale=1]{figs/EQCD_g7_2.pdf}} &= \frac{9}{16} \overline{\gamma}^2 \frac{g_\text{E}^2}{m_\text{E}} N_c d_{abc} d_{abc} \left( \raisebox{-0.55cm}{\includegraphics[scale=1]{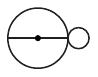}} -2\raisebox{-0.55cm}{\includegraphics[scale=1]{figs/EQCD_spec1.pdf}}+4\raisebox{-0.55cm}{\includegraphics[scale=1]{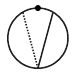}} + \xi \raisebox{-0.55cm}{\includegraphics[scale=1]{figs/EQCD_basketball.pdf}} \right). \label{eq:diagsEQCDg7}
\end{split}
\end{equation}
\end{widetext}
Here, we use a graphical notation for scalar integrals, where the solid (dotted) lines represent massive (massless) propagators with masses normalized to 1, and a black circle indicates that the corresponding propagator is raised to power two; the numerators in all cases are trivial, i.e. just 1. From these expressions, we witness the cancellation of the gauge parameter in $\Delta p^\text{res}_\text{EQCD}$, serving as a valuable check of the correctness of the procedure.

The computation of the loop integrals proceeds with the implementation of integration-by-parts reduction to master integrals \cite{Schroder:2003kb,Schroder:2005va,Kajantie:2003ax}. The relevant linear relations are the following \footnote{We thank York Schröder for discussions on integration-by-parts reductions and three-dimensional vacuum integrals}:
\begin{widetext}
\begin{equation}
\begin{split}
    \raisebox{-0.55cm}{\includegraphics[scale=1]{figs/EQCD_spec1.pdf}} &= \frac{(3d-8)}{6(d-3)} \raisebox{-0.55cm}{\includegraphics[scale=1]{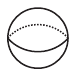}} - \frac{(d-2)}{2(d-3)} \raisebox{-0.55cm}{\includegraphics[scale=1]{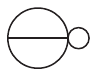}}, \\
    \raisebox{-0.55cm}{\includegraphics[scale=1]{figs/EQCD_spec2.pdf}} &= \frac{(3d-10)(3d-8)}{12(d-4)(d-3)}\raisebox{-0.55cm}{\includegraphics[scale=1]{figs/EQCD_basketball3.pdf}} + \frac{(d-2)^2}{6(d-4)(d-3)}\raisebox{-0.55cm}{\includegraphics[scale=1]{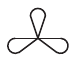}} + \frac{(d-2)(2d-9)}{12(d-3)} \raisebox{-0.55cm}{\includegraphics[scale=1]{figs/EQCD_tadsunset2.pdf}}, \\
    \raisebox{-0.55cm}{\includegraphics[scale=1]{figs/EQCD_spec4.pdf}} &= -\frac{(3d-8)}{4(d-4)} \raisebox{-0.55cm}{\includegraphics[scale=1]{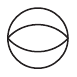}} - \frac{(d-2)^2}{2(d-4)(d-3)} \raisebox{-0.55cm}{\includegraphics[scale=1]{figs/EQCD_tadpoles.pdf}}, \\
    \raisebox{-0.55cm}{\includegraphics[scale=1]{figs/EQCD_mercedes.pdf}} &= \frac{(3d-8)}{12(d-4)} \raisebox{-0.55cm}{\includegraphics[scale=1]{figs/EQCD_basketball2.pdf}} - \frac{(d-2)^2}{6(d-4)(d-3)} \raisebox{-0.55cm}{\includegraphics[scale=1]{figs/EQCD_tadpoles.pdf}} - \frac{2(d-2)}{3(d-4)} \raisebox{-0.55cm}{\includegraphics[scale=1]{figs/EQCD_tadsunset2.pdf}} - \frac{2(d-3)}{3(d-4)} \raisebox{-0.55cm}{\includegraphics[scale=1]{figs/EQCD_spec3.pdf}}, \\
    \raisebox{-0.55cm}{\includegraphics[scale=1]{figs/EQCD_tadsunset.pdf}} &= \frac{1}{3}(3-d) \raisebox{-0.55cm}{\includegraphics[scale=1]{figs/EQCD_tadsunset2.pdf}}.
\end{split}
\end{equation}
With these reductions at hand, $\Delta p^\text{res}_\text{EQCD}$ to $O(\alpha_s^{7/2})$ in Eq.~\eqref{eq:g7res} yields
\begin{equation}
\begin{split}
    \Delta p^\text{res}_\text{EQCD}\biggr\rvert_{\alpha_s^{7/2}} = -\frac{9}{32}\bar{\gamma}^2\frac{g_\text{E}^2}{m_\text{E}}N_c d^{abc}d^{abc}&\left( \frac{2(d-2)(3d-8)}{3(d-4)(d-3)}\raisebox{-0.55cm}{\includegraphics[scale=1]{figs/EQCD_basketball3.pdf}}-\frac{(3d-8)}{3(d-4)}\raisebox{-0.55cm}{\includegraphics[scale=1]{figs/EQCD_basketball2.pdf}}-\frac{d}{(3d-12)}\raisebox{-0.55cm}{\includegraphics[scale=1]{figs/EQCD_spec3.pdf}}\right. \\
    &\left.+\frac{2d(11-7d+d^2)}{3(d-4)(d-3)}\raisebox{-0.55cm}{\includegraphics[scale=1]{figs/EQCD_tadsunset2.pdf}}  \right).
\end{split}
\end{equation}
The analytic results for the master integrals in $d=3-2\varepsilon$ up to the required order in $\varepsilon$ are known from Refs.~\cite{Kajantie:2003ax,Rajantie:1996np}, and read
\begin{equation}
\begin{split}
\label{eq:3dints}
    \raisebox{-0.55cm}{\includegraphics[scale=1]{figs/EQCD_tadsunset2.pdf}} &= \left( \frac{\bar{\Lambda}^{2}_\text{3d}}{4} \right)^{3\varepsilon} \frac{1}{(4\pi)^3}\left[ -\frac{1}{4\varepsilon}-1+\ln \frac{3}{2}+\varepsilon\left(-3-\frac{\pi^2}{48}-2\ln^2 2+2\ln 3 \ln \frac{4}{3}+2\ln \frac{9}{4}-3 \text{Li}_2\left(\frac{1}{3}\right)\right) \right], \\
    \raisebox{-0.55cm}{\includegraphics[scale=1]{figs/EQCD_basketball3.pdf}} &= \left( \frac{\bar{\Lambda}^{2}_\text{3d}}{4} \right)^{3\varepsilon} \frac{1}{(4\pi)^3} \left[ -\frac{3}{4\varepsilon}-6+3\ln\frac{3}{2}+\varepsilon\left( -\frac{17}{16}\pi^2-39-6\ln^2\frac{3}{2}+24\ln\frac{3}{2}-3\text{Li}_2\left(\frac{1}{3}\right) \right) \right], \\
    \raisebox{-0.55cm}{\includegraphics[scale=1]{figs/EQCD_basketball2.pdf}} &= \left( \frac{\bar{\Lambda}^{2}_\text{3d}}{4} \right)^{3\varepsilon} \frac{1}{(4\pi)^3} \left[ -\frac{1}{\varepsilon}-8+4\ln 2-4\varepsilon\left( 13+\frac{17}{48}\pi^2+\ln^2 2-8\ln2 \right) \right], \\
    \raisebox{-0.55cm}{\includegraphics[scale=1]{figs/EQCD_spec3.pdf}} &= \left( \frac{\bar{\Lambda}^{2}_\text{3d}}{4} \right)^{3\varepsilon} \frac{1}{(4\pi)^3} \left( -\frac{\pi^2}{12}+\frac{1}{2}\ln^23+2\text{Li}_2\left( \frac{1}{3} \right) \right),
\end{split}
\end{equation}
\end{widetext}
where $\text{Li}_2(x)$ is the dilogarithm function. We cross-checked the analytic results in Eq.~\eqref{eq:3dints} to high accuracy with an independent fully numerical evaluation using \texttt{FIESTA5} \cite{Smirnov:2021rhf}.

Adding up the different terms yields the following finite, manifestly positive result:
\begin{widetext}
\begin{equation}
\begin{split}
\label{eq:as7/2}
    \Delta p^\text{res}_\text{EQCD}\biggr\rvert_{\alpha_s^{7/2}}=\frac{1}{16}\left(\frac{\alpha_s}{\pi}\right)^{7/2} \frac{\mu_\text{q}^2 T^2N_c N_f^2d^{abc}d^{abc}}{\sqrt{\frac{N_c}{3}+\frac{N_f}{6}\left(1+3\left(\frac{\mu_\text{q}}{\pi T}\right)^2\right)}} \left( 1+\frac{5\pi^2}{12}-\frac{8}{3}\ln 2+\frac{4}{3}\ln 3-\frac{1}{2}\ln^2 3-4\text{Li}_2\left(\frac{1}{3}\right) \right),
\end{split}
\end{equation}
\end{widetext}
Setting $N_c=N_f=3$, we can read off the value of the coefficient $c_3$ reported in Eq.~\eqref{eq:c2c3}.

Note that the $\Delta p^\text{naive}_\text{EQCD}$ terms derived from the diagrams in Eq.~\eqref{eq:g7res} enter at $O(\alpha_s^4)$---beyond the order in which we work here.

\section{$\Delta p$ to higher perturbative orders}
\label{appsec:deltaphighorder}

Here we discuss the extension of our diagrammatic evaluation of $\Delta p$ to higher orders in the coupling expansion, focusing on the potential IR issues and associated effective-theory contributions that may arise along the way \footnote{We thank Guy Moore for discussions on this point.}.

Starting with the $T=0$ case, the finiteness of the diagrams contributing to $\Delta p$ at $O(\alpha_s^4)$ (five-loop) level no longer holds. Indeed, at this order, correcting a gluon line with a quark loop in the Bugblatter diagrams gives rise to a logarithmic IR divergence (in addition to UV divergences):
\begin{equation}
    \raisebox{-0.45\height}{\includegraphics[height=1.6cm]{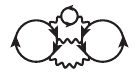}}.
\end{equation}
The IR resummation of this diagram necessitates knowledge of the HTL limit of one specific contribution to the three-loop gluon polarization tensor at $O(\alpha_s^3)$, i.e. two quark loops connected by two gluon lines,
\begin{equation}
    \raisebox{-0.45\height}{\includegraphics[height=1.6cm]{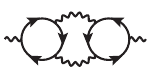}} +
    \raisebox{-0.45\height}{\includegraphics[height=1.6cm]{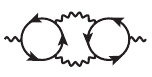}}.
\end{equation}
This computation would allow one to derive the leading nonanalytic term in the perturbative expansion at vanishing temperature, entering at $O(\alpha_s^4\ln\alpha_s)$.

Keeping the temperature nonzero, the sensitivity to the IR begins already at the $O(\alpha_s^3)$ (four-loop) level, as was observed in the computation described in this paper. This resulted in additional $O(\alpha_s^3\ln \alpha_s)$ and $O(\alpha_s^{7/2})$ contributions to $\Delta p$, which were computed consistently within the EQCD effective theory by counting the $C$-odd cubic effective interactions. Going to $O(\alpha_s^4)$ in the naive loop expansion, the IR issues are not only present in a higher number of diagrams, but they also become more severe. Guided by the structure of EQCD, we may further expect diagrams to exhibit power-IR-divergent behavior, resulting in more involved IR subtractions that may compromise the numerical accuracy in this specific kinematic regime. Examples include the following five-loop graphs:
\begin{equation}
    \raisebox{-0.45\height}{\includegraphics[height=1.6cm]{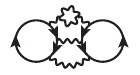}}, \quad
    \raisebox{-0.45\height}{\includegraphics[height=1.6cm]{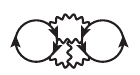}}.
\end{equation}
Notably, from the effective-theory perspective, the contributions to $O(\alpha_s^4\ln \alpha_s)$ and $O(\alpha_s^{9/2})$---available through a massive four- and five-loop vacuum calculation, respectively---are computable with existing tools. This is made possible by the extensive literature on vacuum QFT computations, allowing the determination of the soft contributions to $\Delta p$ to remarkably high order in the coupling expansion.

Finally, at nonzero temperature, we anticipate genuine nonperturbative contributions to $\Delta p$, arising from over-occupied magnetostatic gluon modes of momentum $\alpha_s T$ \cite{Linde:1980ts}. As we argue below, such contributions enter $\Delta p$ at $O(\alpha_s^5)$, at which point lattice simulations of the magnetostatic theory are required to make sense of the weak-coupling expansion.

It is well known that nonperturbative effects originating at the ultrasoft scale $\alpha_s T$ enter the high-temperature pressure of QCD at $O(\alpha_s^3)$, or four loops, as pointed out by Linde \cite{Linde:1980ts}. At leading order, the Linde problem is only present in purely bosonic four-loop QCD diagrams, in particular in the magnetic (i.e., spatial) zero-mode gluonic sector. However, beyond the four-loop level, the Linde problem is also present in diagrams that involve subgraphs with loops propagating heavy modes at the scale $\sim \pi T$, mixing up hard and ultrasoft scales within the same graph. 

Concentrating on the resummed $\Delta p$ quantity defined in Eq.~\eqref{eq:resummedp}, here we aim to identify the leading contribution that satisfies the following two conditions:
\begin{enumerate}
    \item is sensitive to the $d^{abc}$ structure, and therefore survives in the difference $\Delta p$; and
    \item at least one of its corresponding subgraphs is sensitive to the ultrasoft $\alpha_s T$ scale.
\end{enumerate}
Going beyond four-loop level, the number of diagrams in $\Delta p^\text{naive}$ increases rapidly. Therefore, instead of power counting different QCD topologies in $\Delta p^\text{naive}$ one by one (effectively verifying points 1.~and 2.~separately), our strategy is to analyze magnetostatic QCD (MQCD)---the high-temperature effective theory of QCD that captures nonperturbative effects at the ultrasoft scale. Vacuum graphs within this theory automatically satisfy point 2., and thus we just have to work out at what order in the QCD coupling $g$ the $d^{abc}$ structure contributes to the pressure of this theory.

The MQCD theory is defined by the pure-gauge lagrangian
\begin{equation}
    L_\text{MQCD}= \frac{1}{2} \text{tr}\,\bar{F}_{ij}\bar{F}_{ij} + \delta L_\text{MQCD},
\end{equation}
where $\delta L_\text{MQCD}$ are higher-dimensional operators constructed from low-energy spatial gauge fields $\bar{A}_i$ and their derivatives, not relevant for the present argument. The only dimensionful quantity at this order is the magnetic coupling $g_\text{M}$ entering the field-strength tensor $\bar{F}_{ij}$, which encodes the physics at the soft scale $gT$ upon integrating out the massive $\bar{A}_0$ field in EQCD perturbatively.
This means that the magnetic coupling $g_\text{M}$ is obtained perturbatively as a function of the EQCD effective parameters in Eq.~\eqref{appeq:EQCDlagrangian}, $g_\text{M}=g_\text{M}(g_\text{E},m_\text{E},\bar{\gamma},\dots)$. At high enough order in $g$, where nonperturbative effects enter $\Delta p$, we may write, similarly to Eq.~\eqref{eq:resummedp}, the relation
\begin{equation}
\begin{split}
    \Delta p &= (\Delta p^\text{naive}-\Delta p^\text{naive}_\text{EQCD}-\Delta p^\text{naive}_\text{MQCD}) \\
    &\quad+ \Delta p^\text{res}_\text{EQCD} + \Delta p^\text{res}_\text{MQCD}.
\end{split}
\end{equation}
In the low-energy theory, the charge conjugation symmetry is broken by effective operators containing an odd number of $\bar{A}_0$ fields, implying there is no difference between the PQ and QCD magnetostatic theories at the operator level. Therefore, a nonzero contribution to $\Delta p^\text{res}_\text{MQCD}$ can only arise through the dependence of the magnetic coupling on the EQCD matching coefficient $\bar{\gamma}$.

To compute $g_\text{M}$, one performs a matching procedure between EQCD and MQCD. 
In practice, this is most conveniently done in the background-field gauge, where one uses the relation
\begin{align}
    g^2_\text{M} &= \left[1+ \frac{d}{d\mathbf{p}^2}\Pi_\text{T}(0)\right]^{-1}g^2_\text{E} \\
    &= g^2_\text{E} - g^2_\text{E}\frac{d}{d\mathbf{p}^2}\Pi_\text{T}(0) +\dots,
\end{align}
where $\Pi_\text{T}(\mathbf{p}^2)$ is the transverse component of the two-point function for the background field.
With the help of the EQCD Feynman rules and dimensional analysis, the first diagram in the matching involving $\text{tr}\,\bar{A}_0^3$ insertions is parametrically
\begin{equation}
    \frac{d}{d\mathbf{p}^2}\raisebox{-0.55cm}{\includegraphics[scale=1]{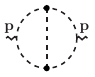}}\biggr\rvert_{\mathbf{p}=0} \sim \frac{g_\text{E}^2}{m_\text{E}^4} \bar{\gamma}^2 T. \label{eq:matching}
\end{equation}
As a result, the difference between the magnetic couplings of the PQ and QCD theories is of order $O(g^6)$,
\begin{equation}
    g_\text{M}^2 \biggr\rvert_{\text{PQ-QCD}} \sim \frac{g_\text{E}^4}{m_\text{E}^4} \bar{\gamma}^2T \sim g^6 \frac{\mu_\text{q}^2T^3}{(\mu_\text{q}^2+T^2)^2}+O(g^8), \label{eq:magcoupling}
\end{equation}
where we suppress numerical prefactors to highlight the parametric dependence on the parameters $T$, $\mu_\text{q}$ and $g$.

By dimensional analysis, the pressure of the MQCD theory is $O(g^6_\text{M}T)\sim O(g^6 T^4)$. Therefore, the leading MQCD contribution to the PQ-QCD difference is
\begin{equation}
    \Delta p^\text{res}_\text{MQCD} \sim Tg_\text{M}^4\times g_\text{M}^2\biggr\rvert_{\text{PQ-QCD}}\sim g^{10}\frac{\mu_\text{q}^2T^6}{(\mu_\text{q}^2+T^2)^2}, 
\label{eq:deltaMQCD}
\end{equation}
i.e. it is $O(\alpha_s^5)$. Remarkably, once the 2-loop matching computation for $g_\text{M}$ is performed, the coefficient of Eq.~\eqref{eq:deltaMQCD} can be known immediately, as the nonperturbative evaluation of the vacuum energy density of MQCD to $O(g_\text{M}^6 T)$ is already known from Ref.~\cite{DiRenzo:2006nh}.

\bibliography{references.bib}

\end{document}